\documentclass{article}

\usepackage[nonatbib, preprint]{neurips_2025}

\usepackage[utf8]{inputenc} 
\usepackage[T1]{fontenc}    
\usepackage{hyperref}       
\usepackage{url}            
\usepackage{booktabs}       
\usepackage{amsfonts}       
\usepackage{nicefrac}       
\usepackage{microtype}      
\usepackage{xcolor}         
\usepackage{marvosym}
\usepackage{amsmath}   
\usepackage{amssymb}  
\usepackage{bm}       
\usepackage{graphicx}
\usepackage{indentfirst}
\usepackage{enumitem}
\usepackage{xcolor}
\usepackage{multirow}
\usepackage{adjustbox} 

\usepackage{graphicx}
\usepackage{subcaption}
\usepackage{caption}

\usepackage{makecell}
\usepackage{pifont}

\newcommand{\cmark}{\ding{51}} 
\newcommand{\xmark}{\ding{55}}

\title{DualMark: Identifying Model and Training Data Origins in Generated Audio}

\author{
Xuefeng Yang\textsuperscript{1, *}
Jian Guan\textsuperscript{1, *, $\dagger$},
Feiyang Xiao\textsuperscript{1, *},
Congyi Fan\textsuperscript{1}, \\
\textbf{Haohe Liu\textsuperscript{2}, Qiaoxi Zhu\textsuperscript{3}, Dongli Xu\textsuperscript{4}, Youtian Lin\textsuperscript{5}}
\\ \\
$^1$Harbin Engineering University  \quad
$^2$University of Surrey \quad
$^3$University of Technology Sydney \\
$^4$LU Leuven \quad
$^5$Nanjing University \\
}

\begin{document}

\footnotetext[1]{Equal contribution}
\footnotetext[2]{Corresponding author}

\maketitle

\begin{abstract}
\label{abs}

Existing watermarking methods for audio generative models only enable model-level attribution, allowing the identification of the originating generation model, but are unable to trace the underlying training dataset. This significant limitation raises critical provenance questions, particularly in scenarios involving copyright and accountability concerns. To bridge this fundamental gap, we introduce DualMark, the first dual-provenance watermarking framework capable of simultaneously encoding two distinct attribution signatures, i.e., model identity and dataset origin, into audio generative models during training. Specifically, we propose a novel Dual Watermark Embedding (DWE) module to seamlessly embed dual watermarks into Mel-spectrogram representations, accompanied by a carefully designed Watermark Consistency Loss (WCL), which ensures reliable extraction of both watermarks from generated audio signals. Moreover, we establish the Dual Attribution Benchmark (DAB), the first robustness evaluation benchmark specifically tailored for joint model-data attribution. Extensive experiments validate that DualMark achieves outstanding attribution accuracy (97.01\% F1-score for model attribution, and 91.51\% AUC for dataset attribution), while maintaining exceptional robustness against aggressive pruning, lossy compression, additive noise, and sampling attacks, conditions that severely compromise prior methods. Our work thus provides a foundational step toward fully accountable audio generative models, significantly enhancing copyright protection and responsibility tracing capabilities.

\end{abstract}

\section{Introduction}
\label{sec:1}

Recent advances in audio generative models have significantly improved the quality of AI-generated audio, enabling realistic music synthesis~\cite{agostinelli2023musiclm, copet2023simple}, immersive auditory environments~\cite{kreuk2023audiogen,liu2023audioldm}, and human-like speech generation~\cite{wang2023neural,vyas2023audiobox,borsos2023audiolm}. While these technological breakthroughs have unlocked numerous innovative applications, they also introduce substantial concerns regarding unauthorized use, copyright infringement, and the ethical implications of deploying audio generative models and Artificial Intelligence Generated Content (AIGC).

\begin{figure}[t!]
    \centering
    \includegraphics[width=0.98\linewidth]{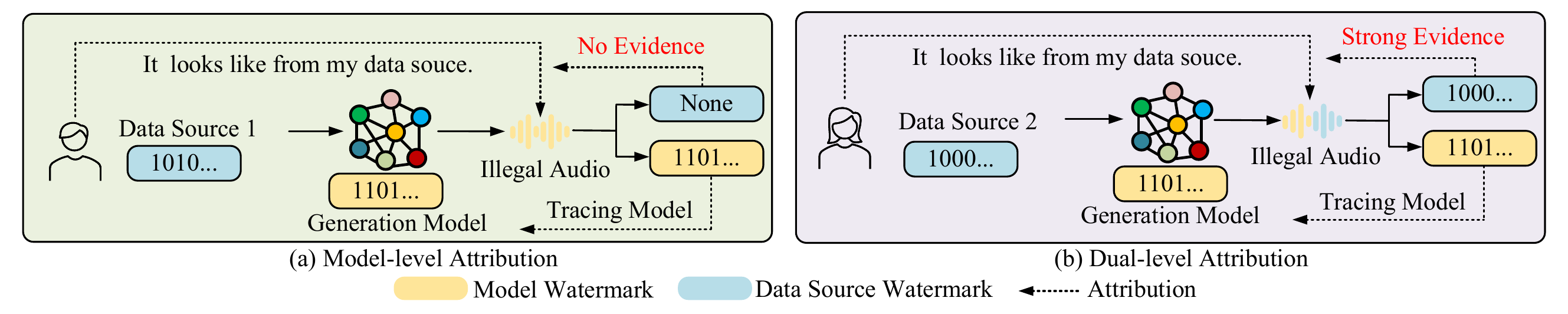}
    \vspace{-2mm}
    \caption{Illustration of differences between the overall pipelines of existing audio watermark attribution methods and our proposed DualMark framework.}
    \label{fig:1}
    \vspace{-5mm}
\end{figure}

Current audio watermarking methods~\cite{cox1996secure, o1996watermarking, cho2022attributable, NEURIPS2024_5d9b7775} primarily focus on model-level attribution, aiming to determine whether a given audio sample was generated by a specific model.
However, as illustrated in Figure~\ref{fig:1}(a), these methods lack the capability to trace the origin of the training data. As a result, they fall short in ensuring accountability and transparency at the data provenance level, thereby failing to protect the rights of data providers.
For instance, watermarking techniques such as AudioSeal~\cite{roman2024proactive} and Timbre~\cite{liu2023detecting} employ external watermark plugins added post-generation, making them highly vulnerable to simple pruning attacks that can remove the watermark without significantly impacting audio quality. Moreover, methods like LatentWM~\cite{san2025latent} integrate watermark embedding directly into model parameters, offering improved robustness against pruning but still failing to preserve detailed information about the training data sources during the generative process. 
Therefore, such approaches leave data creators and providers inadequately protected, increasing the risk of legal disputes and ethical violations.

To address this critical limitation, we propose DualMark, a Dual-Provenance Watermarking framework that enables simultaneous attribution of both the generative audio model and the training dataset.
As illustrated in Figure~\ref{fig:1}(b), DualMark integrates two distinct watermarks, i.e., a model watermark~(marked in yellow) for identifying the specific generative model, and a data watermark~(marked in blue) indicating the origin of the training data.
With these watermarks learning, DualMark can offer strong evidence and safeguard data providers.
To realize this, we introduce a novel Dual Watermark Embedding (DWE) module for embedding both watermarks seamlessly into Mel-spectrogram representations.
Moreover, to ensure robust decoding and watermark integrity throughout the generative process, we propose a dedicated Watermark Consistency Loss (WCL) optimized concurrently with the standard diffusion loss.
To push the boundaries of current evaluation frameworks, we also establish a Dual Attribution Benchmark (DAB), the first comprehensive benchmark explicitly designed to evaluate the robustness and reliability of dual-level watermark attribution under realistic audio processing scenarios.

Extensive experiments demonstrate that DualMark achieves reliable model and data-level attribution, obtaining 97.01\% F1-score for model attribution and 91.51\% AUC for dataset attribution. Importantly, our method outperforms existing post-processing watermark methods under pruning, and remains robustness under typical audio perturbations, e.g., lossy compression, resampling, and noise. Crucially, unlike prior methods, DualMark uniquely enables effective tracing of the training data source, addressing a significant gap in data provenance capabilities. Our contributions thus lay the foundation for more accountable and transparent generative audio technologies, providing robust intellectual property protection and clear provenance assurances for both model developers and data providers.

Our main contributions are summarized as follow:
\begin{itemize}[
    leftmargin=15pt,
    itemsep=0.5ex,
    topsep=0pt,            
    partopsep=0pt,         
    parsep=0pt             
]
    \item We introduce the Dual-Provenance Watermarking Framework (DualMark), the first watermarking framework that enables simultaneous model- and data-level attribution in audio generative models. It fills a critical gap in existing methods that support only model-level identification.
    \item We design a novel Dual Watermark Embedding (DWE) mechanism that embeds two distinct watermark signatures directly into Mel-spectrogram representations during training. One identifies the generation model and the other indicates the training data origin.
    \item We develop a dedicated Watermark Consistency Loss (WCL) that is jointly optimized with the diffusion loss to ensure robust preservation and reliable decoding of both watermark vectors throughout the audio generation process, even under signal degradation.
    \item We establish the Dual Attribution Benchmark (DAB), the first robustness evaluation suite for dual attribution, which assesses performance under real-world audio processing operations, including pruning, lossy compression, resampling, and additive noise. 
\end{itemize}

\section{Related Work}

\subsection{Audio Generation Models}

Recent advances in audio generation models have led to various model architectures, including Generative Adversarial Networks (GANs)~\cite{goodfellow2014generative}, autoregressive models~\cite{borsos2023audiolm, kreuk2023audiogen}, and diffusion models~\cite{liu2023audioldm, liu2024audioldm, liu2024audiolcm}. Among these approaches, diffusion models, particularly AudioLDM~\cite{liu2023audioldm}, have emerged as dominant methods due to their superior audio quality and controllability in latent space. 

Given a Mel-spectrogram $\mathbf{S} \in \mathbb{R}^{T \times F}$, the variational autoencoder (VAE) \cite{kingma2022autoencodingvariationalbayes} encoder $f_{\mathrm{enc}}$ maps it to a latent audio feature $\mathbf{z} = f_{\mathrm{enc}}(\mathbf{S})$, which is then corrupted via a forward diffusion process:
\begin{equation}
    q(\mathbf{x}_t \mid \mathbf{x}_0) = \mathcal{N}\left(\sqrt{\bar{\alpha}_t} \mathbf{x}_0,\; (1 - \bar{\alpha}_t) \mathbf{\epsilon} \right),
\end{equation}
where $\mathbf{x}_0 = \mathbf{z}$, $\mathbf{x}_t$ denotes the corrupted noisy audio feature, $\bar{\alpha}_t$ denotes the noise schedule to control the degree of corruption at the time step $t$, and $\mathbf{\epsilon} \sim \mathcal{N}(\mathbf{0}, \mathbf{I})$ denotes a random noise from a Gaussian distribution. The denoising network $f_\theta(\cdot, \cdot, \cdot)$ is trained to recover the added noise using the latent diffusion loss $\mathcal{L}_{\mathrm{LDM}}$:
\begin{equation}
    \mathcal{L}_{\mathrm{LDM}} = \mathbb{E}_{t, \mathbf{x}_0, \mathbf{\epsilon}} \left\| \mathbf{\epsilon} - f_{\theta}(\mathbf{x}_t, t, \mathbf{e}) \right\|^2_2.
\end{equation}
This latent representation provides a suitable interface for embedding watermark information, motivating our adoption of AudioLDM as the generative backbone.

\subsection{Audio Watermarking Methods for Audio Generation}

Audio watermarking methods embed identifiable information into generated audio signals to ensure intellectual property protection and content attribution. Existing approaches primarily target model-level attribution and are generally classified as external plugin-based and intrinsic parameter-based methods.

External plugin-based methods include post-processing techniques (e.g., WavMark~\cite{chen2023wavmark}, AudioSeal~\cite{roman2024proactive}, and Timbre~\cite{liu2023detecting}), which add watermarks to audio signals after generation, and pre-processing techniques (e.g., GROOT~\cite{liu2024groot}, WMCodec~\cite{zhou2025wmcodec}, and TraceableSpeech~\cite{zhou2024traceablespeech}), which insert watermarks into intermediate representations prior to audio generation. While these methods offer simplicity, their external nature makes them susceptible to pruning attacks, enabling adversaries to remove watermarks without significant audio quality degradation.

Intrinsic parameter-based methods, such as LatentWM~\cite{san2025latent} and HiFiGANw~\cite{HiFi-GANw2024Cheng}, embed watermarking mechanisms directly within the model parameters, enhancing robustness against pruning attacks. However, these methods are limited to model-level attribution, unable to trace back to original training data sources, thus inadequately protecting data providers' rights and lacking essential transparency.

In contrast, our proposed proposed DualMark framework uniquely addresses these limitations by simultaneously embedding both model- and data-level watermarks directly during audio generation, enabling robust and precise attribution of generated audio content back to the originating model and its training data source.

\subsection{Data Provenance and Attribution}

Data provenance and attribution methods have been extensively studied within the visual 
generation domains~\cite{asnani2024promark, zhao2023recipe}, highlighting the importance of tracing training data origins for accountability and intellectual property protection~\cite{agnew2024sound,barnett2024exploring,du2024sok}. However, similar solutions in audio generation are notably absent, resulting in a significant gap regarding reliable data-source attribution. Existing audio watermarking approaches have not addressed this essential aspect, underscoring the need for dedicated methods capable of data-level attribution.

\subsection{Positioning Our Work}

Our work explicitly fills the critical gaps left by existing audio watermarking methods. Our DualMark is the first watermarking framework designed to simultaneously support both model-level and data-level attribution. This comprehensive approach not only enhances robustness against common adversarial perturbations, including pruning, compression, and noise attacks, but also uniquely provides the ability of attribution back to the original training datasets. Furthermore, we introduce the Dual Attribution Benchmark (DAB), a specialized benchmark to  evaluate the effectiveness and robustness of dual-level watermark attribution under realistic conditions, establishing a foundation for future research in accountable and transparent audio generation.

\section{Problem Definition}

We consider the problem of dual-level attribution in audio generative models, where the objective is to identify both the model that generated a given audio signal $\mathbf{s}$ and the training data source that influenced its generation.

Formally, let ${\mathcal{G}_1, \dots, \mathcal{G}_M}$ be a set of $M$ audio generative models and ${\mathcal{C}_1, \dots, \mathcal{C}_N}$ be a set of $N$ training data sources, where each model $\mathcal{G}_i$ is trained on a subset of data sources $\mathcal{C}_j$. Given an audio signal $\mathbf{s}$ synthesized by a model $\mathcal{G}_i$ trained with data from $\mathcal{C}_j$, the goal is to find a mapping as
\begin{equation}
\Psi:\; \mathbf{s} \mapsto (\hat{\mathcal{G}}, \hat{\mathcal{C}}),
\end{equation}
where $\hat{\mathcal{G}}$ denotes the predicted audio generation model, and $\hat{\mathcal{C}}$ indicates the inferred data source responsible for the generation of $\mathbf{s}$.

The attribution is performed by jointly decoding both components using a similarity-based inference function as
\begin{equation}
\Psi(\mathbf{s}) = \left(
\arg\max_{\mathcal{G} \in \mathbb{G}} \kappa(f_{\mathcal{G}}(\mathbf{s}), \mathcal{G}),
\arg\max_{\mathcal{C} \in \mathbb{C}} \kappa(f_{\mathcal{C}}(\mathbf{s}), \mathcal{C})
\right),
\end{equation}
where $f_{\mathcal{G}}(\mathbf{s})$ and $f_{\mathcal{C}}(\mathbf{s})$ are predictions for model- and data-level attribution, and $\kappa$ denotes a similarity function that measures alignment with candidate identifiers. This formulation enables fine-grained, simultaneous tracing of model identity and data provenance from a single audio sample.

\section{Proposed Method}
\label{sec:method}

\begin{figure}[t]
    \centering
    \includegraphics[width=.9\linewidth]{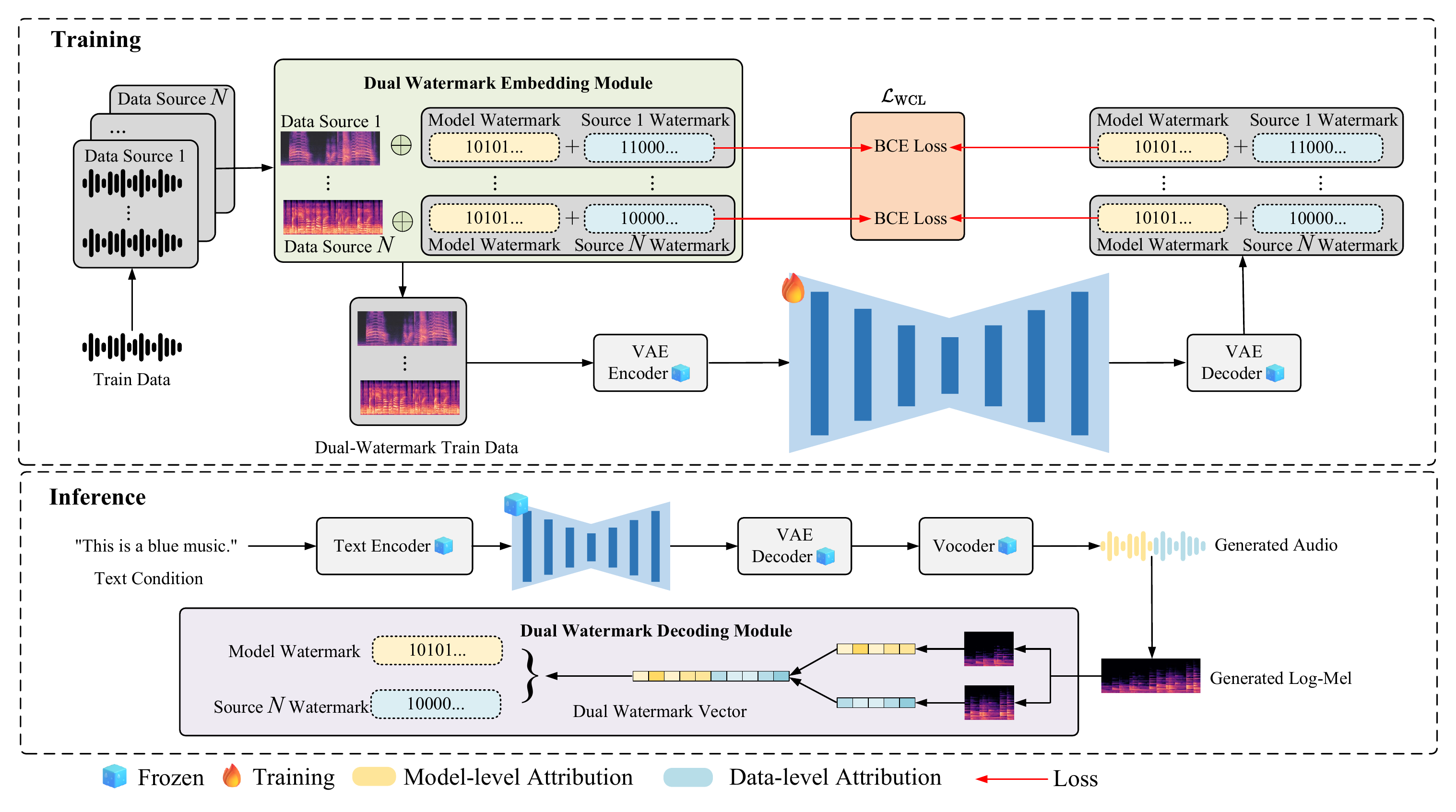}
    \caption{Overview of the proposed DualMark framework. In the training phase, model and data-level watermark vectors are embedded into Mel-spectrograms using a dual watermark embedding (DWE) module. The audio generation model is fine-tuned with the watermark consistency loss (WCL) that enforces watermark consistency. In the inference phase, the generated audio is converted to a Mel-spectrogram and passed through a watermark decoder to recover the dual watermark vector, enabling simultaneous attribution of the generation model and the training data source.}
    \vspace{-4mm}
    \label{fig:pipeline}
\end{figure}

The proposed DualMark is a unified framework for dual-level attribution in audio generation, enabling simultaneous tracing of both the generating model and the original training data. It operates in two stages: a training phase, which embeds model and data watermarks into audio via dual watermark embedding and watermark consistency loss, thereby guiding the model to generate watermarked outputs; and an inference phase, which performs dual watermark decoding from generated audio to achieve  model and data-level attribution. Figure~\ref{fig:pipeline} illustrates the overall architecture.

\subsection{Dual Watermark Embedding}
\label{ssec:dual_watermark_embedding}

In the training phase, we embed the dual watermark vector into the Mel-spectrograms of training samples, to create the watermarked training samples for the following attribution consistency constrained fine-tuning process. The dual watermark vector is a binary encoding vector, composed of both the model watermark vector and the data watermark vector:
\begin{equation}
\label{eq:watermark_message}
    \mathbf{v} = [\mathbf{v}_{\text{M}} \;\| \;\mathbf{v}_{\text{D}}],
\end{equation}
where $\mathbf{v}_{\text{M}} \in \mathbb{F}_2^K$ denotes the model watermark vector of the audio generation model, $\mathbf{v}_{\text{D}} \in \mathbb{F}_2^K$ represents the data watermark vector of an original training data source and $K$ is the dimension of the model and data watermark vectors. The concatenation operation $[\cdot \| \cdot]$ combines them into the dual watermark vector $\mathbf{v} \in \mathbb{F}_2^{I}$, where $I=2K$.

Then, the dual watermark vector is embedded into the Mel-spectrogram $\mathbf{S}$ by a pretrained watermark encoder $\Phi_E(\cdot)$ derived from the pretrained watermarking model RoSteALS~\cite{bui2023rosteals}, as
\begin{equation}
    \widetilde{\mathbf{S}} = \varphi\left(\psi(\mathbf{S}) + \Phi_E(\mathbf{v})\right),
\end{equation}
where $\widetilde{\mathbf{S}}$ denotes watermarked Mel-spectrogram, $\psi(\cdot)$ and $\varphi(\cdot)$ are the encoder and decoder of an adapter autoencoder to map the Mel-spectrogram into the hidden feature space and recover back to the Mel-spectrogram. This resulting watermarked Mel-spectrogram is used in the subsequent model training stage to enable the audio generation model to learn and generate both types of watermark signals during generation.

\subsection{Watermark Consistency Loss}
\label{ssec:fine_tuning}

To enable the model to generate audio with embedded dual watermarks, we fine-tune the latent diffusion backbone (i.e., AudioLDM~\cite{liu2023audioldm}) using a joint objective that balances watermark preservation and audio fidelity.

Each training sample is assigned a dual watermark vector $\mathbf{v}$, which is the combination of both model and data watermark vectors. The watermark encoder $\Phi_E(\cdot)$ injects this watermark vector into the Mel-spectrogram. A corresponding pretrained watermark decoder $\Phi_D(\cdot)$ then predicts the dual watermark vector from the generated spectrogram $\hat{\mathbf{S}}$, as
\begin{equation}
    \hat{\mathbf{v}} = \Phi_D(\hat{\mathbf{S}}),
\end{equation}
where $\hat{\mathbf{v}}$ is the predicted dual watermark vector. Then, the watermark consistency loss $\mathcal{L}_{\text{WCL}}$ is computed using binary cross-entropy:
\begin{equation}
    \mathcal{L}_{\text{WCL}} = -\frac{1}{I} \sum_{i=1}^{I} \left [ \mathbf{v}_i \log \hat{\mathbf{v}}_i + (1 - \mathbf{v}_i) \log(1 - \hat{\mathbf{v}}_i) \right ],
\end{equation}
The final training objective is:
\begin{equation}
    \mathcal{L}_{\text{total}} = \mathcal{L}_{\text{LDM}} + \lambda \cdot \mathcal{L}_{\text{WCL}},
    \label{eq:total_loss}
\end{equation}
where $\mathcal{L}_{\text{LDM}}$ is the standard latent diffusion loss to ensure the quality of the generated spectrogram, and $\lambda = 0.1$ controls the trade-off. 
During training, only the latent diffusion module is updated, while other modules remain fixed. This ensures the model learns to produce high-quality audio with embedded, decodable watermark vectors.

\subsection{Dual Watermark Decoding}

In the inference phase, DualMark performs dual-level attribution by decoding the embedded watermark from generated audio. Given a generated audio sample, we first convert it into a Mel-spectrogram $\bar{\mathbf{S}}$, which is then passed through the pretrained watermark decoder:
\begin{equation}
    \bar{\mathbf{v}} = \Phi_D(\bar{\mathbf{S}}),
\end{equation}
where $\bar{\mathbf{v}}$ is the recovered dual watermark vector. Following the structure defined in Eq.\eqref{eq:watermark_message}, we split $\bar{\mathbf{v}}$ into two parts:
\begin{equation}
    \bar{\mathbf{v}} = [\bar{\mathbf{v}}_{\text{M}} \;\| \;\bar{\mathbf{v}}_{\text{D}}],
\end{equation}
where $\bar{\mathbf{v}}_{\text{M}}, \bar{\mathbf{v}}_{\text{D}} \in \mathbb{F}_2^K$ correspond to the decoded model and data watermark vectors, respectively.

For attribution, these decoded vectors are compared against a predefined watermark codebook. A match determines both the audio generation model and the training data source responsible for producing the content. This dual attribution enables fine-grained provenance tracing for AIGC audio, addressing both model responsibility and data ownership.

Therefore, our DualMark provides an end-to-end watermarking solution that embeds and decodes model and data identifiers through a two-stage training-inference pipeline. By integrating attribution into the generative process itself, it enables the provenance verification across model and data dimensions.

\section{Experiments}

\subsection{Experimental Setup}

\textbf{Dataset:} Our experiments are conducted on the GTZAN music genre dataset \cite{sturm2013gtzan}, which contains 1,000 audio signals equally distributed across 10 genres (e.g., blues, jazz, classical, rock), where the sampling rate of all audio signals is set as 16 kHz, and the mel-bins is set as 64. Each genre is treated as a distinct data origin to support the original training data attribution. During training, every audio sample is embedded with a dual watermark vectors encoding both model and data origin information into the signal, as described in Section~\ref{ssec:dual_watermark_embedding}. To pair audio with text, we generate text prompts using a simple template: "This is a [genre] music piece". 

To evaluate the attribution capability of audio watermarking methods, we construct an evaluation set containing 100 watermarked and 100 non-watermarked generated audio signals per genre, resulting in a total of 2,000 audio signals.

\textbf{Evaluation Metrics:} In our experiments, we evaluate both the effectiveness of the dual attribution mechanism of our DualMark framework and the perceptual quality of the generated audio signals.

To measure attribution performance, we separately measure attribution at the model-level and the data-level. For model-level attribution, following \cite{yu2021artificial}, we use detection accuracy (Det. Acc), Recall and F1 score, to quantify performance, where the detection accuracy reflects the ability to distinguish generated audio from real audio. For attribution to each data origin, we formulate a one-vs-rest binary classification task. Performance is evaluated using the area under the ROC curve (AUC), F1 score and Recall to capture correct identification of data origins.

To evaluate the impact of watermark embedding for the quality of the generated audio, we conduct a human evaluation study for generated audio signals with two subjective metrics, i.e., overall listening quality (OVL) and relevance to prompt (REL). The OVL metric reflects the naturalness and fluency of audio signals, and the REL metric measures how well the generated audio signal aligns with its conditioning text prompt.

\subsection{Attribution Performance}

We first evaluate whether DualMark achieves reliable model- and data-level attribution. To this end, we compare it against state-of-the-art audio watermarking methods, including AudioSeal~\cite{roman2024proactive} and Timbre~\cite{liu2023detecting}. All methods are implemented using two variants of the AudioLDM backbone, i.e., a small-scale model (AudioLDM-S-Full) and a medium-scale model (AudioLDM-M-Full)~\cite{liu2023audioldm}.

\begin{table*}[t]
    \centering
    \caption{Model- and data-level attribution performance of audio generation models with different audio watermarking methods. Here, pruning attacks removes the external watermarking plugin in the audio watermark attribution methods, stemming from the misuse of open-source methods.}
    \label{tab:attribution}
    \begin{adjustbox}{max width=\textwidth}
    \begin{tabular}{ccccccccc}
        \toprule
        \multirow{2}{*}{\textbf{Audio Generation Model}} & \makecell[c]{\multirow{2}{*}{\textbf{Watermarking Method}}} & \multirow{2}{*}{\textbf{Pruning Attack}} &
        \multicolumn{3}{c}{\textbf{Model-Level Attribution}} & 
        \multicolumn{3}{c}{\textbf{Data-Level Attribution}} \\
        \cmidrule(lr){4-6} \cmidrule(lr){7-9} 
        & & & \textbf{Det. Acc} $\uparrow$ & \textbf{F1} $\uparrow$ & \textbf{Recall} $\uparrow$ & 
        \textbf{AUC} $\uparrow$ & \textbf{F1} $\uparrow$ & \textbf{Recall} $\uparrow$ \\
        \midrule
        \multirow{5}{*}{AudioLDM-S-Full} 
          & \multirow{2}{*}{AudioSeal~\cite{roman2024proactive}} & \xmark & \textbf{97.35} & \textbf{97.28} & \textbf{94.70} & --     & --     & --     \\
          &                                                      & \cmark & 50.00 & 00.00 & 00.00 & --     & --     & --     \\
          \cmidrule(lr){2-9}
          & \multirow{2}{*}{Timbre~\cite{liu2023detecting}} & \xmark & 95.05 & 94.08 & 90.20 & --     & --     & --     \\
          &                                                    & \cmark & 50.00 & 00.20 & 00.10 & --     & --     & --     \\
          \cmidrule(lr){2-9}
          & \textbf{DualMark (Ours)} & -- & \underline{97.10} & \underline{97.01} & \underline{94.20} & \textbf{91.51} & \textbf{93.68} & \textbf{83.60} \\
        \midrule
        \multirow{5}{*}{AudioLDM-M-Full} 
          & \multirow{2}{*}{AudioSeal~\cite{roman2024proactive}} & \xmark & \textbf{93.40} & \textbf{92.93} & \textbf{86.80} & --     & --     & --     \\
          &                                                      & \cmark & 50.00 & 00.00 & 00.00 & --     & --     & --     \\
          \cmidrule(lr){2-9}
          & \multirow{2}{*}{Timbre~\cite{liu2023detecting}} & \xmark & 90.90 & 90.00 & 81.90 & --     & --     & --     \\
          &                                                    & \cmark & 50.10 & 00.60 & 00.30 & --     & --     & --     \\
          \cmidrule(lr){2-9}
          & \textbf{DualMark (Ours)} & -- & \underline{92.65} & \underline{92.07} & \underline{85.30} & \textbf{87.55} & \textbf{90.83} & \textbf{75.78}   \\
        \bottomrule
    \end{tabular}
\end{adjustbox}
\vspace{-4mm}
\end{table*}

\begin{figure}[t!]
  \centering
  \includegraphics[width=0.85\linewidth]{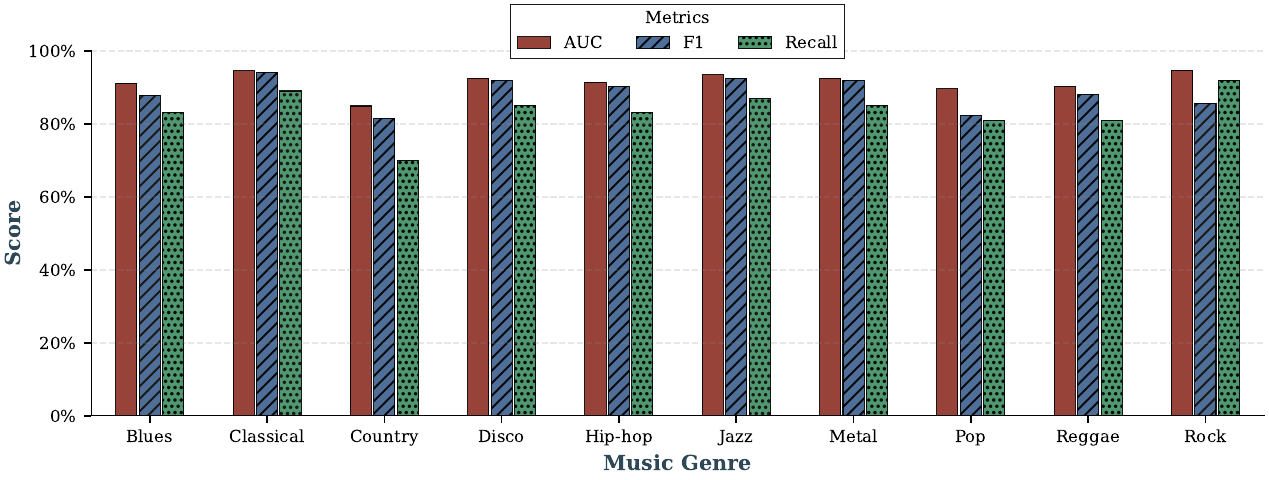}
  \caption{Performance for data-level attribution across multiple original training data origins.}
  \vspace{-4mm}
  \label{fig:data-acc}
\end{figure}

Table~\ref{tab:attribution} reports the comparative results. For model-level attribution, the proposed DualMark achieves comparable performance with existing methods, validating its ability to reliably identify generated audio and help prevent unauthorized use of audio generation models. For data-level attribution, the proposed DualMark can effectively achieve data attribution, while AudioSeal and Timbre do not support attribution to the training data origin at all. This highlights the unique capability of our DualMark framework to trace both the audio generation model and its underlying training data, filling a critical gap left unaddressed by the compared methods.

In addition, we also evaluate robustness under the pruning attack that removes any external watermarking plugins. As shown in Table~\ref{tab:attribution}, model attribution accuracy for AudioSeal and Timbre collapses, whereas DualMark is unaffected because its watermarking capability is integrated into the model parameters. Therefore our DualMark ensures reliable attribution for responsible AI.

We further report per-genre performance for data-level attribution in Figure~\ref{fig:data-acc}. Across all genres, DualMark achieves performance in terms of AUC, F1 and Recall metrics, above 80\% in all cases except the Recall of the genre country. These results demonstrate that the embedded watermark vectors are effectively preserved during generation, overcoming the degradation of the watermark information faced by prior methods and enabling effective data-level attribution. In summary, DualMark provides an effective solution for protecting audio content through the dual-level attribution, contributing to responsible use and oversight of AI-generated audio.

\subsection{Ablation Study}

We further conduct an ablation study to evaluate the impact of our watermark consistency loss $\mathcal{L}_{\text{WCL}}$ on enabling DualMark to achieve both model-level and data-level attribution. As described in Section~\ref{ssec:fine_tuning}, this loss is designed to preserve the content of embedded watermark vectors during audio generation, making it central to the achieving dual attribution.

\begin{table}[htbp]
  \centering
  \caption{Ablation study to explore the impact of our watermark consistency loss $\mathcal{L}_{\text{WCL}}$ on model- and data-level attribution performance.}
  \label{tab:ablation_watermark}
  \begin{tabular}{ccccccc}
    \toprule
    \multirow{2}{*}{\textbf{Using $\mathcal{L}_{\text{WCL}}$}} & \multicolumn{3}{c}{\textbf{Model-Level Attribution}} & \multicolumn{3}{c}{\textbf{Data-Level Attribution}} \\
    \cmidrule(lr){2-4} \cmidrule(lr){5-7}
     & \textbf{Det. Acc} $\uparrow$ & \textbf{F1} $\uparrow$ & \textbf{Recall} $\uparrow$ & \textbf{AUC} $\uparrow$ & \textbf{F1} $\uparrow$ & \textbf{Recall} $\uparrow$ \\
    \midrule
    \xmark & 50.00 & 0.00 & 0.00 & 49.98 & 47.44 & 0.10 \\
    \cmark & \textbf{97.10} & \textbf{97.01} & \textbf{94.20} & \textbf{91.51} & \textbf{93.68} & \textbf{83.60} \\
    \bottomrule
  \end{tabular}
\end{table}

We compare DualMark variants fine-tuned with and without the watermark consistency loss $\mathcal{L}_{\text{WCL}}$ using the AudioLDM-S-Full backbone. As shown in Table~\ref{tab:ablation_watermark}, removing this loss leads to a dramatical drop in attribution performance. For model-level attribution, both F1 score and Recall fall to zero, while the detection accuracy falls into 50\%, equivalent to random guessing. For data-level attribution, Recall drops near zero, while AUC and F1 score fall below 50\%, equivalent to random guessing yet. This sharp decline confirms that the watermark consistency loss is crucial for preserving watermark information during audio generation.

\begin{figure}[h]
    \vspace{-3mm}
    \centering
    \includegraphics[width=0.8\linewidth]{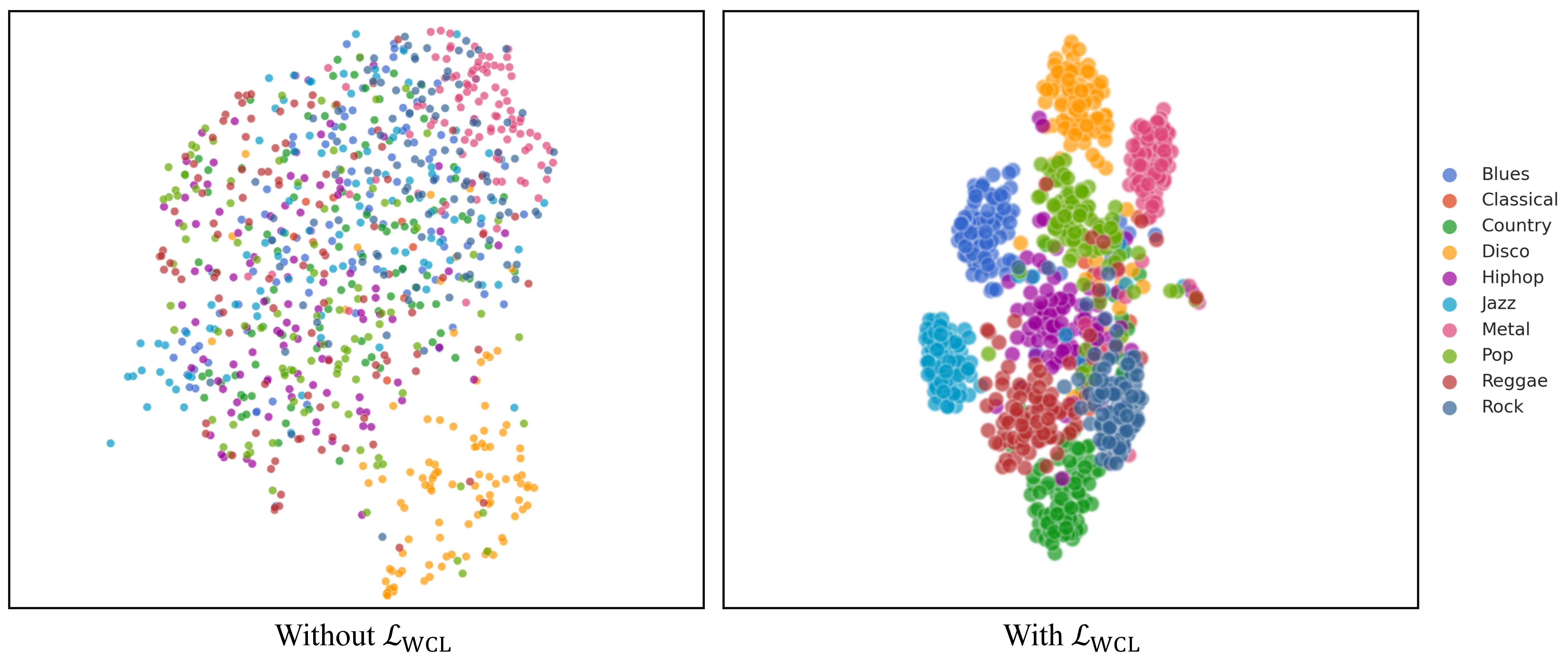}
    \caption{The t-SNE visualization of the latent features of generated audio, without and with the watermark consistency loss $\mathcal{L}_{\text{WCL}}$.}
    \label{fig:tsne}
     \vspace{-3mm}
\end{figure}

To further illustrate the effect of this loss, we visualize the latent features of generated audio using t-SNE, with and without $\mathcal{L}_{\text{WCL}}$, as shown in Figure~\ref{fig:tsne}. Without the loss, the latent features are entangled and lack clear separation across training data sources, making it difficult to extract unique watermark vectors for tracing the training data origin. In contrast, with $\mathcal{L}_{\text{WCL}}$, features from the same data origin cluster tightly, while those from different origins are clearly separated. This structured representation enables accurate extraction of origin-specific data watermark vectors to trace the training data origin, supporting effective data-level attribution.

These results highlight the key role of the watermark consistency loss in preserve watermark details for reliable traceability to both the audio generation model and its training data, reinforcing the importance of this component in building responsible and accountable AIGC systems.

\subsection{Robustness Evaluation}

To measure the dual attribution robustness of the proposed DualMark framework, we evaluate its performance under various audio signal processing operations applied to the generated audio signals. Specifically, we consider four common types of audio signal processing operations, i.e., white/pink noise addition, resampling, and AAC lossy compression, each of which may occur in real-world audio transmission or editing pipelines.

\begin{table}[htbp]
  \centering
  \caption{Dual attribution performance under different types of audio signal processing operations for generated audio signals with our DualMark.}
  \label{tab:robustness_ours}
  \setlength{\tabcolsep}{3.1pt}
  \begin{small}
  \begin{tabular}{cccccccc}
    \toprule
    \multirow{2}{*}{\textbf{Attack Type}} & \multirow{2}{*}{\textbf{Operation}} & 
    \multicolumn{3}{c}{\textbf{Model-Level Attribution}} &
    \multicolumn{3}{c}{\textbf{Data-Level Attribution}} \\
    \cmidrule(lr){3-5} \cmidrule(lr){6-8} 
    & & \textbf{Det. Acc} $\uparrow$ & \textbf{F1} $\uparrow$ & \textbf{Recall} $\uparrow$ & \textbf{AUC} $\uparrow$ & \textbf{F1} $\uparrow$ & \textbf{Recall} $\uparrow$ \\
    \midrule
    --                  & Clean Audio & \textbf{97.10} & \textbf{97.01} & 94.20 & \textbf{91.51} & \textbf{93.68} & \textbf{83.60} \\
    \multirow{2}{*}{\textbf{Noise Attack}} & 
     Add White Noise          & 89.00 & 87.64 & 78.00 & 79.50 & 85.05 & 59.50 \\
     &  Add Pink Noise           & 92.60 & 92.01 & 85.20 & 84.75 & 89.17 & 70.00 \\
    \textbf{Sampling Attack} & Resample            & \textbf{97.10} & 94.20 & \textbf{97.01} & 90.75 & 93.35 & 82.00 \\
    \textbf{Lossy Compression Attack} & AAC Compression      & 96.40 & 96.27 & 92.80 & 91.25 & 93.67 & 83.00 \\
    \bottomrule
  \end{tabular}
  \end{small}
\end{table}

Table~\ref{tab:robustness_ours} shows that DualMark withstands all tested perturbations. Model-level attribution stays strong, with F1 and Recall above 78.00\% in every case and surpassing 92.80\% after resampling or AAC compression. Data-level attribution is similarly stable, e.g., AUC remains above 79.50\% and F1 above 85.0\% across all operations. Remarkably, model-level Recall peaks at 97.01\% under resampling, outperforming even the clean-audio baseline. These findings confirm that the embedded watermark vectors with our DualMark framework survive common signal degradations, enabling reliable dual attribution after typical real-world audio processing operations.

\subsection{Evaluation of Audio Quality}

To assess the perceptual impact of embedded watermarks, we conduct a human evaluation study across 4 conditions, i.e., the original backbone, AudioSeal \cite{roman2024proactive}, Timbre \cite{liu2023detecting}, and our DualMark. 16 listeners rated each audio signal on two 1-to-100 scales, i.e., overall listening quality (OVL) and relevance to the text prompt (REL). 
 \vspace{-5mm}

\begin{table}[htbp]
  \centering
  \caption{Subjective evaluation on generated audio signals from AudioLDM backbones with different audio watermarking methods in terms of OVL and REL metrics.}
  \label{tab:subjective-audioldm}
  \begin{tabular}{c l cc}
    \toprule
    Audio Generation Model     & Watermarking Method   & OVL $\uparrow$  & REL $\uparrow$  \\
    \midrule
    Ground Truth               & —                     & 77.09           & 79.57           \\
    \midrule
    \multirow{4}{*}{AudioLDM-S-Full}
                               & Original              & 62.19           & 65.52           \\
                               & AudioSeal \cite{roman2024proactive}            & 63.89           & 65.65           \\
                               & Timbre \cite{liu2023detecting}                & 59.48           & 62.88           \\
                               & \textbf{DualMark (Ours)}      & 60.68           & 63.55           \\
    \midrule
    \multirow{4}{*}{AudioLDM-M-Full}
                               & Original              & 64.23           & 68.82           \\
                               & AudioSeal \cite{roman2024proactive}            & 67.70           & 69.30           \\
                               & Timbre \cite{liu2023detecting}                & 63.34           & 65.85           \\
                               & \textbf{DualMark (Ours)}      & 60.17           & 63.39           \\
    \bottomrule
  \end{tabular}
    \vspace{-2mm}
\end{table}

\begin{figure}
    \centering
    \includegraphics[width=.9\linewidth]{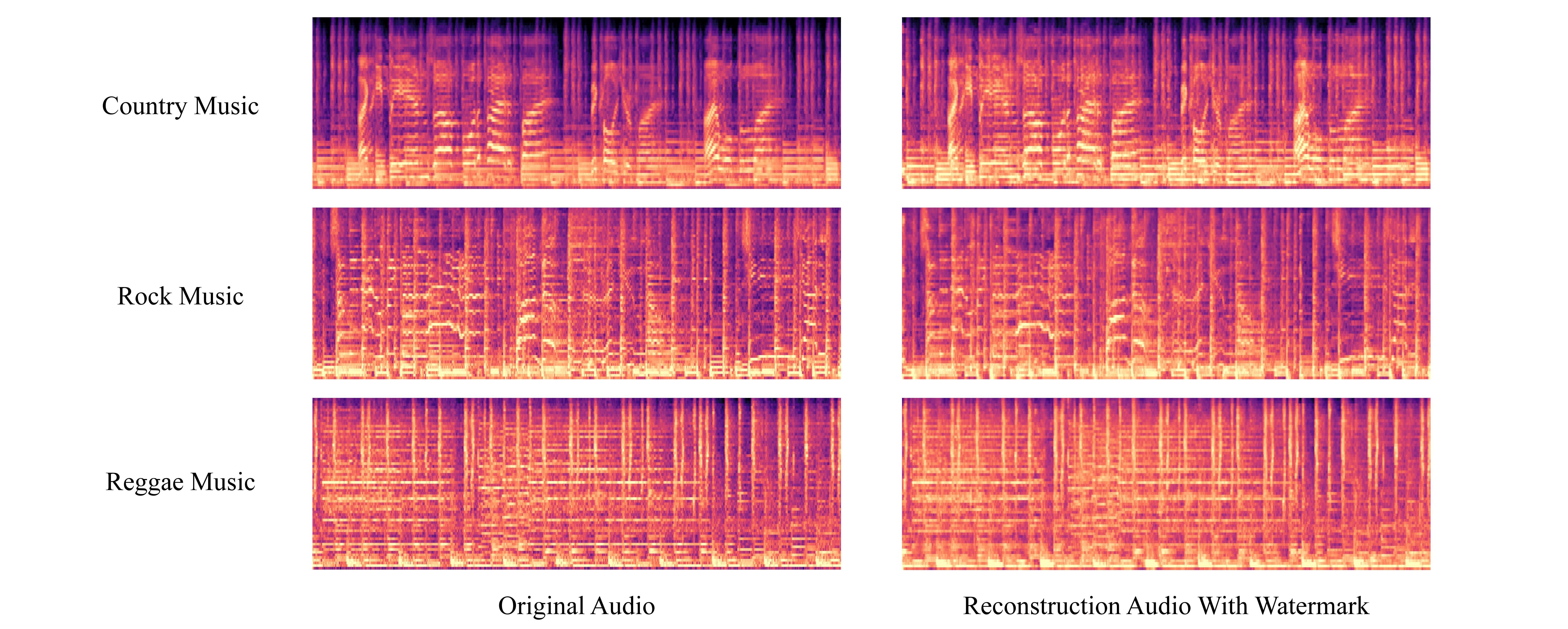}
    \caption{Illustration of Mel-spectrograms of original audio signals and the reconstructed audio signals by the audio generation model with our DualMark.}
    \label{fig:mel_illustration}
     \vspace{-2mm}
\end{figure}

As Table \ref{tab:subjective-audioldm} shows, OVL and REL scores of our DualMark are statistically indistinguishable from the original backbone and from the other watermarking methods, confirming that the dual watermark embedding module and watermark consistency loss do not degrade perceived quality a lot.

To illustrate this further, Figure~\ref{fig:mel_illustration} compares Mel-spectrograms of original audio signals with those reconstructed by the audio generation model with our DualMark framework. The close visual match indicates that spectral details are faithfully preserved. In summary, this verifies that, our DualMark achieves dual attribution while leaving audio fidelity and prompt relevance intact.

\section{Conclusion}

We introduced DualMark, the first framework for dual-provenance attribution in audio generation, enabling simultaneous tracing of both the generation model and training data source from a single audio sample. By embedding binary watermark vectors into Mel-spectrograms and preserving them through a consistency loss, DualMark enables reliable model and data attribution at inference. To support this task, we proposed the Dual Attribution Benchmark (DAB) to evaluate watermark robustness under common signal-level perturbations. While results are promising, DualMark marks an initial step toward traceable audio generation. Future work will explore higher-capacity encoding, vocoder-aware or waveform-level embedding, and expanded evaluation covering open-set, cross-model, and semantic-preserving attacks. We hope this work lays the groundwork for accountable and auditable AIGC systems.

{\small
\bibliographystyle{ieee_fullname}
\bibliography{refs}
}

\appendix
\newpage

\section{Supplementary Experiments}
\subsection{Hyperparameter Selection for watermark consistency loss}

To further understand the effect of the weight of the watermark consistency loss on attribution performance, we vary the hyperparameter $\lambda$ in the joint training objective defined in Equation~\eqref{eq:total_loss}. This hyperparameter weight controls the relative importance of the watermark consistency loss with respect to the primary diffusion loss.

Table~\ref{tab:wm_strength_attribution} summarizes model-level and data-level attribution performance under different values of $\lambda$. We observe that attribution performance improves consistently as $\lambda$ increases from 0.01 to 0.10, reaching peak accuracy at $\lambda = 0.10$. However, further increasing $\lambda$ leads to a noticeable decline in both model- and data-level attribution metrics. This suggests that overly strong emphasis on watermark supervision may interfere with the learning of audio generation quality or distributional robustness.

These results highlight the importance of carefully balancing the diffusion and watermark objectives, and empirically validate the choice of $\lambda = 0.10$ as an effective setting for DualMark.

\begin{table}[h]
    \centering
    \caption{Impact of hyperparameter weight $\lambda$ on model-level and data-level attribution performance.}
    \label{tab:wm_strength_attribution}
    \begin{tabular}{ccccccc}
        \toprule
        \multirow{2}{*}{$\lambda$} &
        \multicolumn{3}{c}{\textbf{Model-Level Attribution}} &
        \multicolumn{3}{c}{\textbf{Data-Level Attribution}} \\
        \cmidrule(lr){2-4} \cmidrule(lr){5-7}
        & \textbf{Det. Acc} $\uparrow$ & \textbf{F1} $\uparrow$ & \textbf{Recall} $\uparrow$ &
        \textbf{AUC} $\uparrow$ & \textbf{F1} $\uparrow$ & \textbf{Recall} $\uparrow$ \\
        \midrule
        0.01 & 94.55 & 94.24 & 89.10 & 88.08 & 91.10 & 76.89 \\
        0.05 & 95.80 & 95.62 & 91.60 & 89.37 & 92.09 & 79.40 \\
        0.10 & \textbf{97.10} & \textbf{97.01} & \textbf{94.20} & \textbf{91.51} & \textbf{93.68} & \textbf{83.60} \\
        0.20 & 91.95 & 91.25 & 83.90 & 84.43 & 87.72 & 70.00 \\
        0.30 & 88.45 & 86.94 & 76.90 & 87.49 & 87.49 & 64.90 \\
        0.40 & 82.05 & 78.12 & 64.10 & 75.04 & 81.14 & 50.60 \\
        \bottomrule
    \end{tabular}
\end{table}

\subsection{Impact of Different DDIM Steps}

As described in prior work~\cite{liu2023audioldm, san2025latent}, the number of denoising diffusion implicit model (DDIM) steps can influence the performance of audio generation. In this section, we investigate how the number of DDIM steps affects dual attribution performance within the DualMark framework.

The results are presented in Table~\ref{tab:sampling_steps_attribution}. We observe that increasing the number of steps from 100 to 200 improves both model-level and data-level attribution. However, performance slightly declines when the step count increases from 200 to 250. These findings suggest that 200 DDIM steps provide the best empirical balance for maintaining watermark information and maximizing attribution performance in DualMark.

\begin{table}[htb]
    \centering
    \caption{Impact of number of DDIM steps on model-level and data-level attribution performance.}
    \label{tab:sampling_steps_attribution}
    \begin{tabular}{ccccccc}
        \toprule
        \multirow{2}{*}{DDIM Steps} &
        \multicolumn{3}{c}{\textbf{Model-Level Attribution}} &
        \multicolumn{3}{c}{\textbf{Data-Level Attribution}} \\
        \cmidrule(lr){2-4} \cmidrule(lr){5-7}
        & \textbf{Det. Acc} $\uparrow$ & \textbf{F1} $\uparrow$ & \textbf{Recall} $\uparrow$ &
        \textbf{AUC} $\uparrow$ & \textbf{F1} $\uparrow$ & \textbf{Recall} $\uparrow$ \\
        \midrule
        100 steps & 91.95 & 91.25 & 83.90 & 87,33 & 91.16 & 75.10 \\
        150 steps & 91.55 & 90.77 & 83.10 & 85.44 & 89.83 & 71.30 \\
        \textbf{200 steps} & \textbf{97.10} & \textbf{97.01} & \textbf{94.20} & \textbf{91.51} & \textbf{93.68} & \textbf{83.60} \\
        250 steps & 90.20 & 89.14 & 80.40 & 85.73 & 90.00 & 71.90 \\
        \bottomrule
    \end{tabular}
\end{table}

\subsection{Impact of Waveform Reconstruction}
\label{app:with vocoder}

\begin{table}[t]
  \centering
  \caption{Attribution performance for AudioLDM models before and after vocoder reconstruction.}
  \label{tab:audio_generation_attribution}
  \resizebox{\textwidth}{!}{
  \begin{tabular}{l l ccc ccc}
    \toprule
    \multirow{2}{*}{\textbf{Audio Generation Model}} & \multirow{2}{*}{\textbf{Stage}} &
    \multicolumn{3}{c}{\textbf{Model-Level Attribution}} &
    \multicolumn{3}{c}{\textbf{Data-Level Attribution}} \\
    \cmidrule(lr){3-5} \cmidrule(lr){6-8}
    & & \textbf{Det. Acc} $\uparrow$ & \textbf{F1} $\uparrow$ & \textbf{Recall} $\uparrow$ &
          \textbf{AUC} $\uparrow$ & \textbf{F1} $\uparrow$ & \textbf{Recall} $\uparrow$ \\
    \midrule
    \multirow{2}{*}{AudioLDM-S-Full}
      & After Vocoder   & 97.10           & 97.01        & 94.20         & 91.51      & 93.68     & 83.60        \\
      & Before Vocoder  & \textbf{100.00} & \textbf{100.00} & \textbf{100.00} & \textbf{94.48} & \textbf{95.02} & \textbf{89.80} \\
    \midrule
    \multirow{2}{*}{AudioLDM-M-Full}
      & After Vocoder   & 93.80           & 93.39        & 87.60         & 87.55      & 90.83     & 75.78        \\
      & Before Vocoder  & \textbf{100.00} & \textbf{100.00} & \textbf{100.00} & \textbf{94.52} & \textbf{94.81} & \textbf{90.00} \\
    \bottomrule
  \end{tabular}
  }
\end{table}

As discussed above, the waveform reconstruction process can degrade watermark information in generated audio signals. In this section, we investigate the impact of this process, which is handled by the vocoder~\cite{kong2020hifi} module in the AudioLDM backbone. Specifically, we compare attribution performance on Mel-spectrograms directly generated by the model (before vocoder) and on Mel-spectrograms extracted from the vocoder-reconstructed audio (after vocoder).

The results are presented in Table~\ref{tab:audio_generation_attribution}. We observe that attribution performance is consistently higher before waveform reconstruction, with model-level metrics reaching 100\% across the board. This indicates that the vocoder introduces some degradation to the embedded watermark information, albeit to a limited extent. These findings suggest that improving the watermark preservation through the waveform reconstruction stage may be a valuable direction for further enhancing the DualMark framework.

\section{Implementation Details}
\label{ref:ImplementationDetails}

We fine-tune the AudioLDM backbone in the proposed DualMark using a single NVIDIA RTX 3090 GPU with a batch size of 2 for 30 epochs. Training is performed with the Adam optimizer \cite{kingma2014adam} at a learning rate of $1 \times 10^{-4}$, using a linear warm-up over the first 2,000 steps. During training, only the UNet module within the AudioLDM backbone is updated. The watermark consistency loss is applied with a hyperparameter $\lambda$ to balance the dual attribution ability for protection and the quality of  generated audio signals, and we empirically find that the value of 0.1 yields the best trade-off between attribution ability and audio quality. This setting allows the proposed DualMark to achieve high accuracy on both model-level and data-level attribution tasks.

\section{Limitations}
\label{ref:limitations}
While DualMark introduces the first solution for dual-provenance attribution in audio generation and establishes a new benchmark (DAB) for this task, several limitations remain, reflecting the early stage of this research direction. 
\textbf{1) Fixed-Length Binary Encoding:} 
Each identity is represented by a 7-bit binary code, which limits scalability in representing large numbers of models or data sources, and reduces error-correction capacity under adversarial perturbations.
\textbf{2) Vocoder-Agnostic Design:} 
The vocoder is not jointly optimized for watermark preservation, and its decoding process may weaken watermark signals, especially under compression or reconstruction noise.
\textbf{3) Task Scope and Generalization:} 
Our experiments are limited to music-style generation with AudioLDM on GTZAN. Whether DualMark generalizes to speech, environmental sounds, or other backbones (e.g., AudioLM, MusicGen) remains open.
\textbf{4) Benchmark Scope:} 
While we propose the first Dual Attribution Benchmark, it currently focuses on signal-level perturbations and pruning. Future extensions could incorporate semantic-preserving attacks, cross-model transfer, or open-set attribution.

\end{document}